\documentclass[journal=jpcbfk,manuscript=article]{achemso}

\usepackage[version=3]{mhchem} % Formula subscripts using \ce{}
\usepackage{subfig,xcolor,textcomp}

\author{Rodrigo Seraide}
\author{Mario A. Bernal}
\email{mbernalrod@gmail.com}
\author{Gustavo Brunetto}
\affiliation[Universidade Estadual de Campinas]{Instituto de F\'isica Gleb Wataghin. Universidade Estadual de Campinas. SP. Brazil.}
\author{Umberto de Giovannini}
\affiliation[Max Planck Institute]{Max Planck Institute for the Structure and Dynamics of Matter and Center for Free-Electron Laser Science \& Department of Physics. Luruper Chaussee 149. 22761 Hamburg. Germany.}

\alsoaffiliation[Universit\`a degli Studi di Palermo]{Dipartimento di Fisica e Chimica. Universit\`a degli Studi di Palermo. Via Archirafi 36. I-90123. Palermo. Italy.}
\author{Angel Rubio}
\affiliation[Max Planck Institute]{Max Planck Institute for the Structure and Dynamics of Matter and Center for Free-Electron Laser Science \& Department of Physics. Luruper Chaussee 149. 22761 Hamburg. Germany.}
\alsoaffiliation[Nano-Bio Spectroscopy Group]{Nano-Bio Spectroscopy Group and ETSF,  Dpto. F\'isica de Materiales, Universidad del Pa\'is Vasco UPV/EHU, 20018 San Sebasti\'an, Spain.}

\title[proton-DNA collision]{A TDDFT-based Study on the Proton-DNA Collision}

\begin{document}

%\begin{tocentry}
%\end{tocentry}

\begin{abstract}
  The interaction of heavy charged particles with DNA is of interest for several areas, from hadrontherapy to aero-space industry. In this paper, a TD-DFT study on the interaction of a 4 keV proton with an isolated DNA base pair was carried out. Ehrenfest dynamics was used to study the evolution of the system during and after the proton impact up to about 193 fs.  This time was long enough to observe the dissociation of the target, which occurs between 80-100 fs. The effect of base pair linking to the DNA double helix was emulated by fixing the four O3' atoms responsible for the attachment. The base pair tends to dissociate into its main components, namely the phosphate groups, sugars and nitrogenous bases. A central impact with energy transfer of 17.9 eV only produces base damage while keeping the backbone intact. An impact on a phosphate group with energy transfer of about 60 eV leads to backbone break at that site together with base damage, while the opposite backbone site integrity is kept is this situation. As the whole system is perturbed during such a collision, no atom remains passive. These results suggest that base damage accompanies all backbone breaks since hydrogen bonds that keep bases together are much weaker that those between the other components of the DNA.

\end{abstract}


\section{Introduction}

The interaction of ionizing particles with DNA is a very complex process which depends both on the particle track structure (radiation quality) and the genetic material geometrical  conformation. The early physicochemical damage that ionizing radiation induces in DNA may lead to biological effects. These effects are of supreme importance for medical radiation applications, both for diagnostic and therapeutic procedures. In addition,  aerospace industry is interested on this problem as astronauts are exposed to charged particle radiation during their missions, and this includes heavy particles with mass even larger than a proton. The radiobiological problem consists in studying biological effects induced by ionizing particles in living beings. Several approaches have been used to deal with this problem during the past seven decades. {\it In vitro assays}, in which cellular cultures are irradiated and later analyzed, is the main source of information for understanding this problem. This is the case of the pioneer works of Karl Sax and coworkers  \cite{Sax38}, which were used by Lea and Catchside \cite{Lea42} as an empirical base to formulate a successful biophysical model for the early DNA damage. Later, Kellerer and Rossi proposed the long-standing Dual Radiation Action Theory \cite{Kellerer72}, which states that lethal lesions induced by ionizing radiation in cells are produced by the interaction of two sub-lesions (probably double strand breaks, DSB). 

With the rapid increase of computing power along the last few decades,  numerical approaches came out. For instance, Monte Carlo simulation of particle transport can be  combined with a DNA geometrical models and biophysical model in such a way that DNA damage probability can be estimated \cite{Pomplun91,Nikjoo99,Friedland11b,Bernal11b}. The latter approach counts a DNA damage, typical a single strand break (SSB), when an energy deposition above a certain threshold value occurs inside the target in question. Commonly, this target is the sugar-phosphate group. This method implicitly assumes that the collision of the ionizing particle with DNA is a one-body problem. That is, the rest of the DNA molecule remains frozen when the incoming particle interact with the atom in question. This is not the case in reality, mainly when dealing with relatively slow ions which produce a strong perturbation of the target system. 

Time Dependent Density Functional Theory (TD-DFT) emerges as a powerful tool to study the full dynamics  of  collisions involving complex systems since it is capable to account for the  many body problem in a consistent way. First of all, the ground state of the target system is determined using the Density Functional Theory (DFT). According to the Hohenberg-Kohn theorem,  the ground-state density is enough to determine the ground state of an electronic system \cite{Hohenberg64}.  Kohn and Sham \cite{Kohn65} found a way to uncouple the Schrodinger equation system  of the electronic  system, making the problem easier to solve. That is, the Kohn-Sham formalism is able to exactly map the interacting system into a non-interacting and easier to solve problem.  

In principle, TD-DFT can be used to study the collision between a charged particle and DNA or some of its constituents. Bacchus et al. \cite{Bacchus12} studied the collision of carbon ions on nitrogenous bases thymine, uracil and 5-halouracil. Targets were bombarded at various incidence direction and impact parameters. Calculations were carried out with the MOLPRO package \cite{MOLPRO}.  They determined charge transfer cross sections for different carbon ions charge states by following an impact parameter approximation. The authors speculate about dissociation cross sections but they did not study this process directly. Sadr-Arani et al. \cite{Sadr12,Sadr13,Sadr14,Sadr15} have carried out several works using experimental and theoretical methods for studying the fragmentation of DNA/RNA bases such as  uracil, cytosine, adenine, and guanine. Their calculations were based on the DFT formalism by they did not explicitly simulate any collision process. Instead, they stretched bonds up to break and determined the involved dissociation energies and possible fragments.  Lopez-Tarifa et al. \cite{Lopez14} have recently used the TD-DFT approach to study the fragmentation of doubly ionized uracil in gas phase. They did not account for the explicit incidence of any projectile. They simply removed electrons from inner-shells {\it ad hoc} and let the excited molecule evolve in time.

Classical molecular dynamics has been also used to study the collision of charged particles with DNA.  Albofath et al. \cite{Abolfath11} used the reactive force field ReaxFF \cite{vanDuin01} to study the role of hydroxyl  free radicals on DNA damage. They  randomly distributed free hydroxyl radicals in small pockets around a DNA fragment and followed the evolution of the system. They found that OH radicals produce holes in the sugar-moiety rings and that they evolve to larger holes that comprise several  bases. Then this damage propagates to the bases and lead to single and double strand breaks. One year later, Abolfath et al. \cite{Abolfath13} continued studying the same process using  the GEANT4-DNA Monte Carlo package \cite{Bernal15a} to obtain the initial position of hydroxyl radicals. Then, the interaction of those radicals with DNA was described by the REAXFF-based molecular dynamics approach. Primary 1 MeV electrons and protons were studied in this work. They reported that  protons produce four times more DNA double strand breaks that electrons with the same energy. Bottl\"ander et al. \cite{Bottlander15} used the REAX force field provided by Abolfath et al. \cite{Abolfath13} to study the interaction of protons with DNA in a NaCl aqueous solution. They simulated the direct interaction of a proton with a DNA fiber fragment by uniformly distributing the energy transferred by the projectile to the target atoms  within a cylinder with 2 $\AA$ radius. This energy was determined from the particle stopping power. The authors reported the number of SSB and DSB produced by different energy transferred to the medium when the projectile travels along the three main cartesian axes. The effect of a violent sock wave created by ions with a very high stopping power (or linear energy transfer) has been also studied using classical molecular dynamics \cite{Surdutovich13,Vera16}. They used the CHARMM potential model to simulate the evolution of DNA atoms after the passage of the ion so explicit bond breakage was not accounted for. Instead, they estimated energy changes in DNA bonds due the influence of the ion-induced shock wave and speculated on the possible creation of single strand breaks.  Recently, Bacchus-Montabonel and Calvo \cite{Bacchus15,Bacchus16} studied the effect of the hydration shell  around biomolecular targets (uracil and aminooxazole) on the proton-induced charge transfer process. This effect was done by adding  only two water molecules  at different molecular sites. They determined charge transfer cross sections during the impact of 10 eV to 10 keV protons using a software package based on the impact parameter approximation, rather than using TD-DFT calculations.

This work aims at the study of the proton-DNA collision problem using the TD-DFT to see how a base-pair evolves during and after the impact of an energetic proton. This approach should allow the observation of many-body effects during this collision. In addition, in-vacuum dissociation times and the energy required for this dissociation can be estimated  under different conditions, including different impact parameters and bounding with neighbor base-pairs. It should be remarked that the detailed study of the DNA dissociation is out of the scope of this work. We simply want to have a qualitative picture of this process as a support for the introduction of a new approach to study the early DNA damage induced by ionizing radiation. This new method would be an alternative to current biophysical models (discussed above). That is, those approaches based on the assumption that only the atom targeted by the incoming particle is affected while the others remain frozen and that double strand breaks can be induced after the production of two close enough single strand breaks. Up to our knowledge, this is the first time the TD-DFT approach is used to explicitly study the collision between a heavy charged particle and a DNA base pair.

Atomic units are used throughout this work, unless otherwise stated. 

\section{Methods}
\subsection{Theoretical background of the TD-DFT.}

The electronic Schr\"odinger equation of the interacting system with N electrons with positions at $({\bf r_1},{\bf r_2},...,{\bf r_N})$, respectively, is

\begin{equation}
\Big \{-{1\over 2}\sum_{i=1}^N\nabla_i^2+{1\over 2}\sum_{i,j=1}^2 {1\over |{\bf r}_i - {\bf r}_j |}+\sum_{i=1}^N v_{ext({\bf r}_i)} \Big \} \Psi({\bf r_1},{\bf r_2},...,{\bf r_N})=E\Psi({\bf r_1},{\bf r_2},...,{\bf r_N}),
\label{eq1}
\end{equation}

where the first term is the kinetic energy of electrons and the second one is the so-called Hartree term. The external potential  in absence of electromagnetic fields is

\begin{equation}
 v_{ext({\bf r}_i)}=-\sum_{k=1}^M{Z_k\over |{\bf r}_i-{\bf R}_K|}
 \label{eq2}
\end{equation}

and comes from the interaction of electrons with point-like nuclei. After solving equation \eqref{eq1}, the electronic density $n({\bf r})$ can be determined as

\begin{equation}
%n({\bf r})=N\int d^3{\bf r}\Psi({\bf r},{\bf r_2},...,{\bf r_N})
n({\bf r})=N\int d^3r_2\cdots d^3r_N\Psi({\bf r},{\bf r_2},...,{\bf r_N})\Psi^*({\bf r},{\bf r_2},...,{\bf r_N}).
 \label{eq3}
\end{equation}

Equation \eqref{eq1} is very hard to solve but Kohn and Sham found a simpler and exact way to solve it by introducing the exchange-correlation potential  $v_{xc}({\bf r})$. According to their approach, the equation system \eqref{eq1} can be decomposed in equations  for the orbitals $\phi_i({\bf r})$ forming a single Slater determinant of a fictitious non-interacting system with the same density of the interacting one as following 
\begin{equation}
\Big \{-{1\over 2}\nabla^2+v_{Hartree}[n]({\bf r})+ v_{ext}({\bf r})+ v_{xc}[n]({\bf r}) \Big \} \phi_i^{KS}({\bf r})=E\phi_i^{KS}({\bf r}),
\label{eqKH1}
\end{equation}

where

\begin{equation}
v_{Hartree}[n]({\bf r})=\int d^3{\bf r'}{n({\bf r'})\over |{\bf r}-{\bf r'}|}
\label{Hartree}
\end{equation}

and $v_{xc}[n]({\bf r})$ is the exchange-correlation potential which accounts for the many-body effects of the problem. Notice that the Hartree and exchange-correlation potentials are functional of the density defined as

\begin{equation}
n({\bf r})=\sum_{i=1}^N |\phi_i^{KS}({\bf r})|^2.
 \label{KHdensity}
\end{equation}

For the time-dependent case, Kohn-Sham equations are
\begin{equation}
\Big \{ -{1\over 2} \nabla^2+v_{Hartree}[n]({\bf r},t)+ v_{ext}({\bf r},t)+ v_{xc}[n]({\bf r},t) \Big \} \phi_i^{KS}({\bf r},t)=i{\partial \over \partial t}\phi_i^{KS}({\bf r},t),
\label{eqKH2}
\end{equation}

Now both the density $n({\bf r},t)$ and nuclei positions $R_K({\bf r},t)$ are functions of time. Similar to the ground state calculation, equation system \eqref{eqKH2} is solved in a self-consistent way for each time step. Here, we used the Adiabatic Local Density Approximation for describing the time-dependent exchange-correlation functional $v_{xc}[n]({\bf r},t)$ \cite{Marques12}.
The time-evolution of nuclei was described through the Ehrenfest dynamics. In this formalism, nuclei are treated classically and allowed to move under the influence of the mean field generated by electrons. That is, their equation of motion is
\begin{equation}
m_K {\partial^2 {\bf R}_K\over\partial t^2 }=-\nabla_K V({\bf\bar R}),
\label{Ehrenfest}
\end{equation}

where $m_K$ and ${\bf R}_k$ are the nucleus mass and position, respectively. $V({\bf\bar R})$ accounts for the electron-nucleus attraction and nucleus-nucleus repulsion and is a function of the nuclei positions ${\bf\bar R}=({\bf R}_1,{\bf R}_2,...,{\bf R}_M)$. In this approximation the solution to the time dependent Schr\"odinger equation is  obtained propagating the classical equation of motion for the ions \eqref{Ehrenfest} together with the quantum mechanical TDDFT equations for the electrons \eqref{eqKH2} until a given time.

\subsection{Collision setup}

\begin{figure}[h!]
\begin{center}
\includegraphics[width=0.7\textwidth]{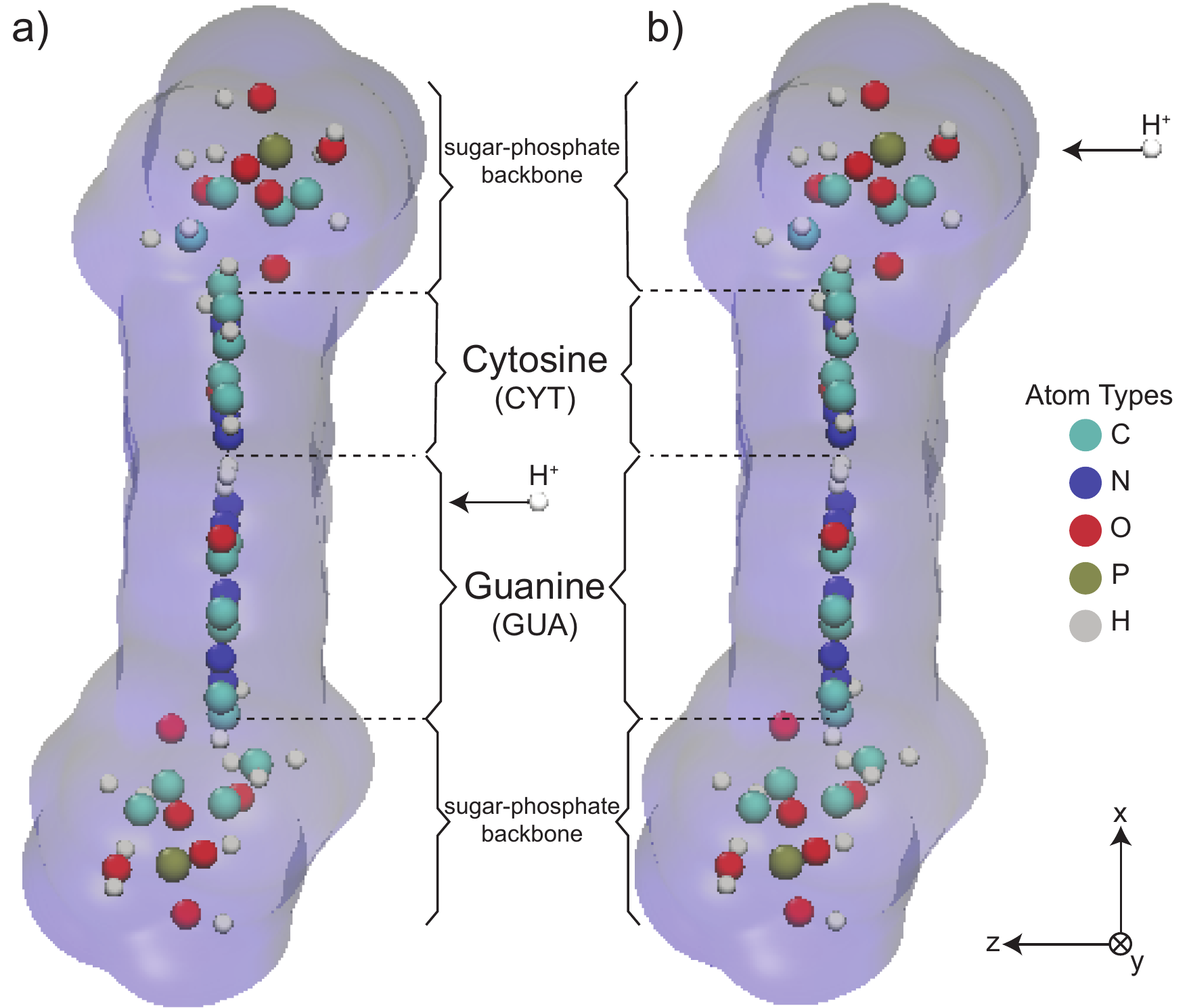}
\caption{Proton-DNA collision setups. }
\label{fig1}
\end{center}
\end{figure}

A proton with about 4 keV energy impacts on an isolated Guanine-Cytosine B-DNA base-pair (bp) at rest.  Atoms positions correspond to canonical B-DNA as that found in Protein Data Bank ID 309D \cite{PDB}. This bp contains the whole phosphate group on one side, ending in O3', and the O3' atom belonging to the phosphate adjacent group. In other words, this bp's backbone ends in two O3' atoms.  These terminal oxygen atoms were fixed in some calculations to simulate the effect of bounding to the adjacent base pairs (see details below). In order to stabilize the molecule, we completed the dangling bonds with hydrogens. This means that hydrogen atoms were added to both O3' terminals and another one was attached to the O2P atom, which is responsible for some DNA-protein binding. Two impact parameters were included in this study. First, the proton impinges the DNA bp with 0 impact parameter with respect to the molecule's geometrical center, near the hydrogen-bridge bonds that link nitrogenous bases. Second, the proton impacts the bp with 0 impact parameter with respect to the upper phosphorus atom, which belongs a the sugar-phosphate group. The proton initially travels along the Z axis, which normally crosses the plane containing the nitrogenous bases atoms. Fig \eqref{fig1} shows the localization of the DNA atoms and the incoming proton. It should be remarked that this proton is treated as any other ion of the target system, as described in eq. \ref{Ehrenfest}.

\subsection{Ground state calculation}

The Octopus code version 4.1.2 \cite{Castro06,Octopus12,Andrade15} was used to carry out TDDFT calculations.  A ground state calculation for the DNA bp was done in a first stage. The Local Density Approximation (LDA) and the modified Perdew\& Zunger LDA \cite{Perdew81} were used for the exchange and correlation functionals, respectively. The system in question was placed inside a 46x20x20  a.u.$^3$ box and the calculation grid was obtained using a 0.4 a.u. spacing along the three Cartesian axes. This spacing was found after an optimization process during which the system total energy converged. Troullier-Martins pseudopotentials were used for all the atoms that conform the bp  \cite{Troullier91} in such a way that their K-shell electrons  were not treated explicitly.  Using these pseudopotentials, the number of orbitals  for the whole bp was 119. 
 
%Self-interaction correction was not accounted for in these calculations.

\subsection{Time-dependent calculations}

 After having obtained the ground state of the DNA bp, the proton was placed {\it ad hoc} according to the impact parameter in question. For the proton with zero impact parameter with respect to the geometrical center of the bp, the initial position was (0,0,-15) a.u. For the proton with zero impact parameter w.r.t. the upper phosphorus atom (Cytosine side), the initial point was (16.721,-4.396,-15) a.u. The initial proton velocity was (0,0,0.4) a.u. in both cases so the proton impinges  normal to the plane defined by the nitrogenous bases. A first calculation was done for the central impact case in which all atoms were  free to move. In a second stage, the four O3' terminal atoms were fixed. This emulates the case in which the bp is bound to a double helix DNA chain.  Time step was set to 0.05 a.u. and the total calculation time was 8000 a.u. ($\sim$193 fs). This time was chosen in such a way that the initial dissociation process can be observed. Calculations were carried out in a 120-core cluster and a single 193 fs calculation took about 6 weeks of wall-clock time.  Absorbing boundary conditions were employed to prevent artificial reflections of electrons at the boundary of the simulation box \cite{Giovannini15}. The complex potential method was used. According to preliminary calculations, a temperature of 300 K shows negligible effects on the evolution of the bond lengths in question. Time-evolution rendering was done with the VMD software  \cite{Humphrey96}.

\section{Results}
\begin{figure}[h]
\includegraphics[width=0.6\textwidth]{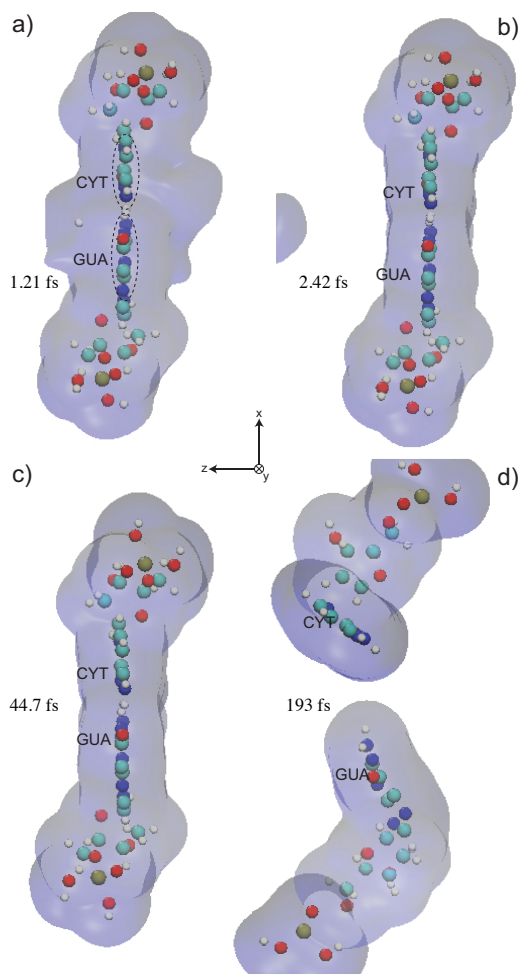} 
\caption{System snapshots for some important stages during the proton-free DNA central collision:  a) the proton just passing through the bp, b) the proton leaves the bp taking a fraction of the system charge (electron capture process), c) beginning of the dissociation process, and d) the bp dissociates with into five fragments. Animation with the whole process can be found in supplementary information.}
\label{diss-free-0}
\end{figure}

\begin{figure}[ht]
\subfloat[]{\label{BLa}\includegraphics[width=0.33\linewidth]{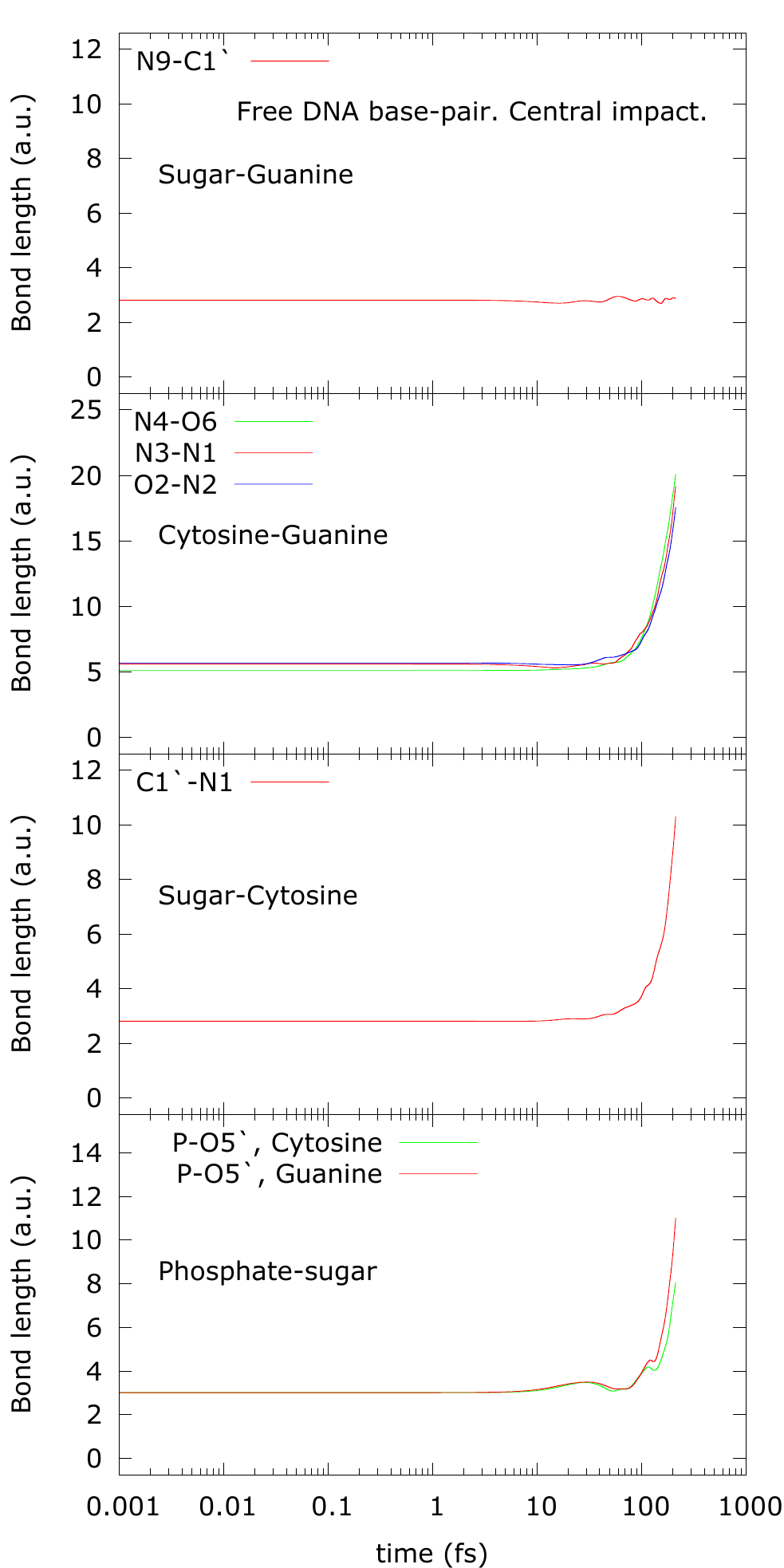}} 
\subfloat[] {\label{BLb}  \includegraphics[width=0.33\linewidth]{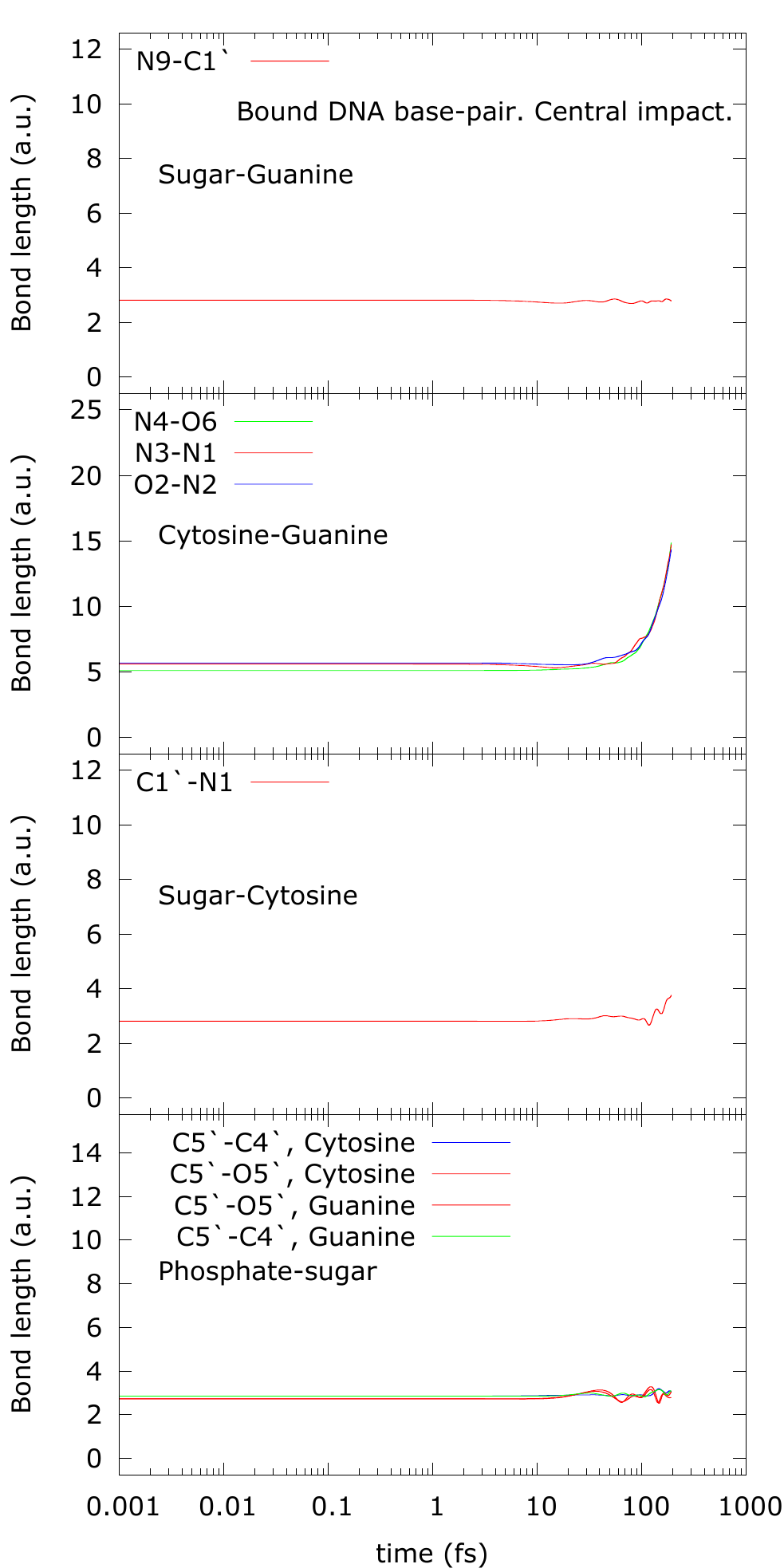}}
\subfloat[]{\label{BLc}\includegraphics[width=0.33\linewidth]{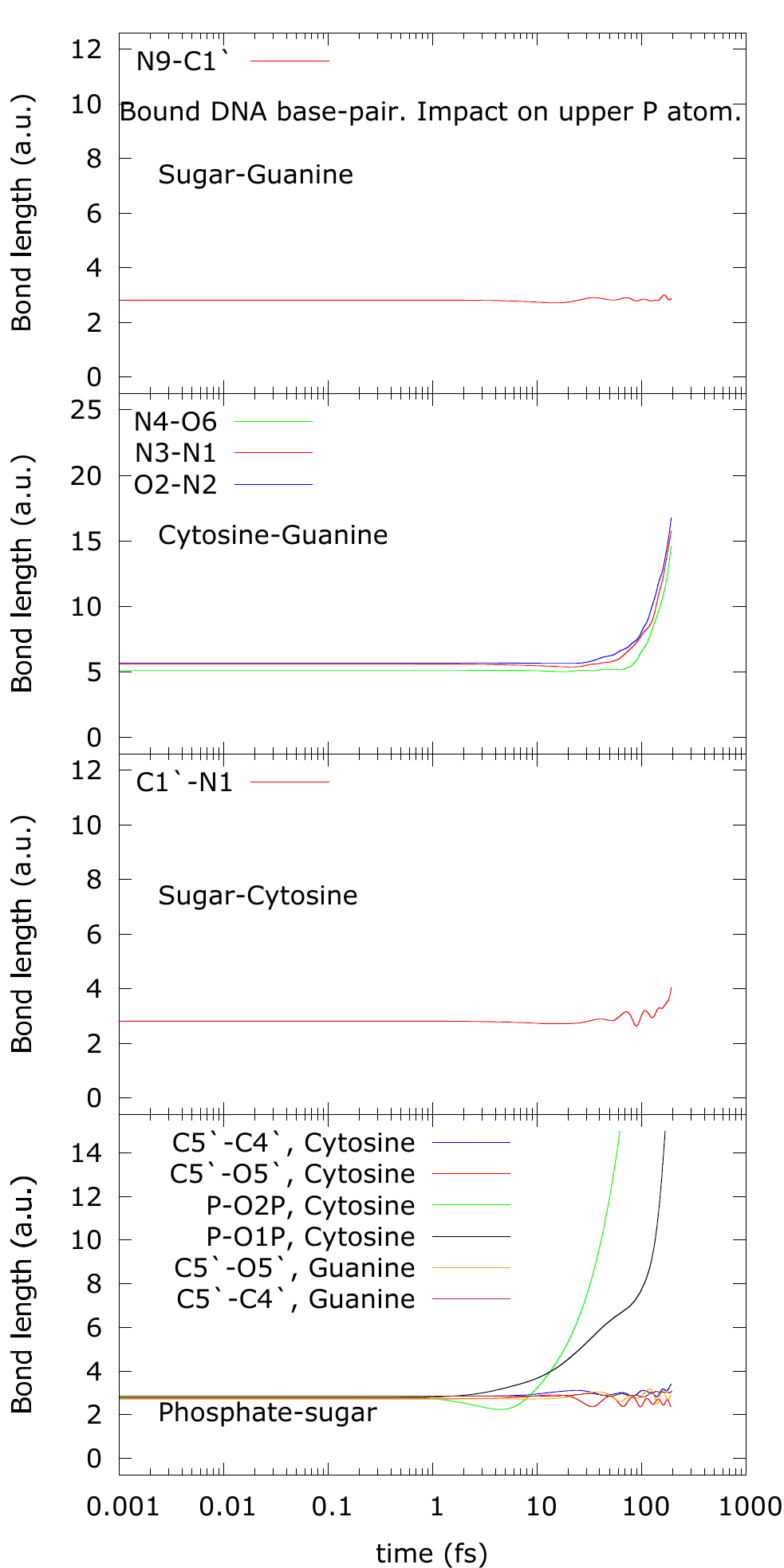}}
\caption{Length of some important DNA bonds as a function of time during and after a collision with a 4 keV proton for cases in which a) all atoms are free and the proton impacts on the bp center, b) all atoms are free except four oxygen atoms that would link the bp to their neighbors and the proton impacts on the bp center, and all atoms are free except four oxygen atoms that would link the bp to their neighbors and c) the proton impacts on one of the phosphorus atoms.}
\label{BL}
\end{figure}

Figure  \ref{diss-free-0} shows snapshots at characteristic times during the evolution of the DNA bp after the proton impact at zero impact parameter where all atoms are free to move. Blue shaded area represents  the 0.001 \% electronic density isosurface to show the evolution during the collision. Snapshot a) captures the proton just passing through the bp.  Snapshot  b) displays the proton charge capture. In the following lines, an energetic analysis will be carried out as a consistency test of our calculations. Yet, it does not aim at a rigorous explanation of the DNA dissociation. The energy transferred by the proton to the bp was 17.9 eV in this collision.  This energy is mainly transferred to the electrons during the collision  and about ~33 \% (6.02 eV) of it is subsequently transferred to the ions until just before the dissociation process start.  At first glance, it seems that the bp tends to dissociate into its main components, namely the phosphate groups, sugars, and bases. However, Fig. \ref{BLa} shows that both O5' atoms dissociate from the corresponding phosphate groups. In fact, they remain attached to the corresponding sugar through the C5' atom (ester bond), which means that the integrity of the deoxyribose sugar prevails over that of the phosphate group. P-O5' (P-O(C)) and P-O2 (P=O) bonds require about 3.67 eV and 6.03 eV for breaking, respectively \cite{Range04}. Thus it is more energetically advantageous to break the P-O5' bond instead of the P-O2 one. It was also obtained that the P-O5' bond length is $\sim$1.593 \AA, which remains stable until  the dissociation process takes place. This value agrees with results reported in the literature (1.591-1.603 \AA) \cite{Range04}. Then, about 7.34 eV would be used to break both P-O5' bonds. The hydrogen-bonds that keep bases together  are relatively weak.  The binding energy of the N-H-N  and N-H-O bonds  are $\sim$0.135 eV and $\sim$0.301 eV, respectively \cite{Legon87}.  In the CG bp, there are  one N-H-N and two N-H-O bonds so the total binding energy for these hydrogen-bonds is around 0.737 eV.  Thus, an energy of about 8.077 eV is used to dissociate the three  hydrogen bonds and two sugar-phosphate bonds so still remain additional 9.823 eV from the transferred energy, which include the kinetic energy transferred to the ions (6.02 eV).  Figure \ref{BLa} shows that the sugar-cytosine bond is broken unlike that between the sugar and the guanine base, which remains stable during the calculation time. C-N bond energy is about 3.158 eV \cite{Huheey93} so it is estimated from these results that  about 11.235 eV  are used to dissociate the DNA bp. The remaining transferred energy should be converted into kinetic energy of the dissociation fragments.

Fig. \ref{BLa} shows  that the DNA bp actually dissociates into five products. Bond breaking occurs so that the main constituents of the DNA separate from each others. That is, the molecule tends to produce fragments such as phosphite groups,  nitrogenous bases and deoxyriboses. This fragmentation pathway seems to be plausible since the fragments produced are relatively stable radicals and the linkage between them should be the weakest bonds of the molecule. However, unlike the sugar-cytosine bond, the sugar-guanine bond is stable until 193 fs.  Sugar-cytosine and cytosine-guanine bonds begin to dissociate almost simultaneously (from $\sim$50 fs on), while sugar-phosphate bonds take a bit longer (from $\sim$80 fs). These dissociation times are consistent with those reported for large molecules ($\sim$0.1 ps) \cite{Gross11}. The time elapsed between the proton impact and the beginning of dissociation was estimated as $\sim$49 fs. These results shows that passage of the proton perturbs the whole system. That is, this is a many-body collision in which no component of the system remains frozen. 

\begin{figure}[h]
\includegraphics[width=0.6\textwidth]{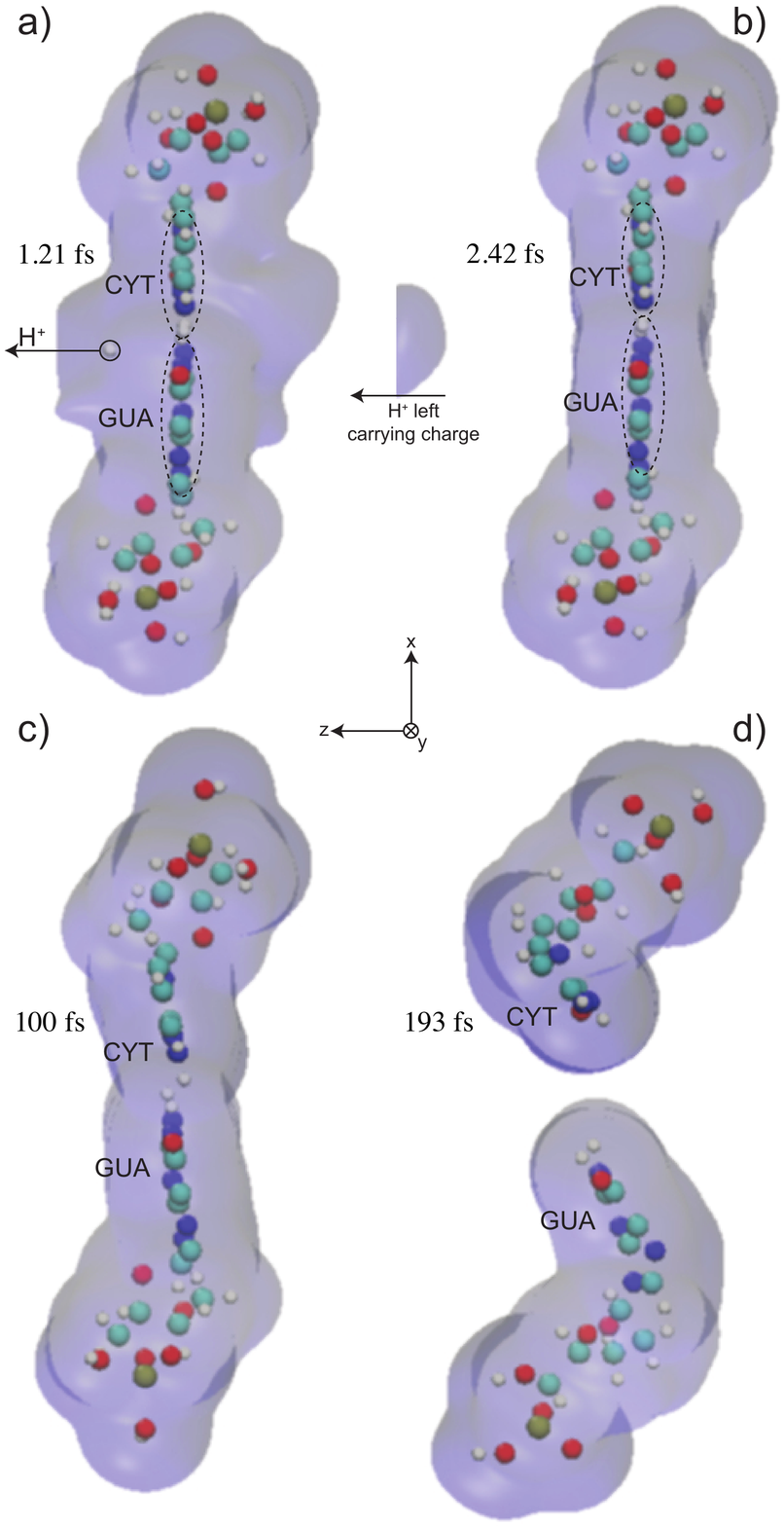}
\caption{System snapshots for some important stages during the proton-bound DNA central collision:  a) the proton just passing through the bp, b) the proton leaves the bp taking a fraction of the system charge (electron capture process), c) beginning of the dissociation process, and d) the bp dissociates with into three fragments. Animation with the whole process can be found in supplementary information.}
\label{diss-bound-0}
\end{figure}

Figure \ref{diss-bound-0} shows the same results as in Fig. \ref{diss-free-0}, but with the  O3' terminal atoms held fixed in order to emulate the binding of the bp to their neighbors. The energy transfer is 17.9 eV  again which means that the binding of the bp to their neighbors does not influence this quantity, at least for central impact. This energy transfer occurs in a very short time so it is possible that the effect of bp linking through oxygen atoms relatively far from the impact region is very weak. As in the free DNA case, about 6.01 eV of this energy is later transferred to the ions after the collision. The dissociation process can be better observed in Fig. \ref{BLb}. The dissociation of the hydrogen bonds  begins almost at the same time as in the free bp case but now the process seems to be slower. At the maximum calculation time, hydrogen bonds lengths are larger for the bound bp than for the free one.  Sugar-cytosine bond seems to be in dissociation route but at a slower velocity, beginning at $\sim$100 fs and thus delayed compared with the free configuration. Unlike the free bp situation, phosphate-sugar bonds do not dissociate. This means that only the so-called base damage occurs while the DNA backbone is not broken. The cytosine base is ejected  while  guanine  remains attached to the corresponding sugar.  That is, only three fragments are produced in this case. This behavior would be expected as the bp is now linked to their neighbors and the impact was on the hydrogen bonds that keep bases together.

\begin{figure}
\includegraphics[width=0.6\textwidth]{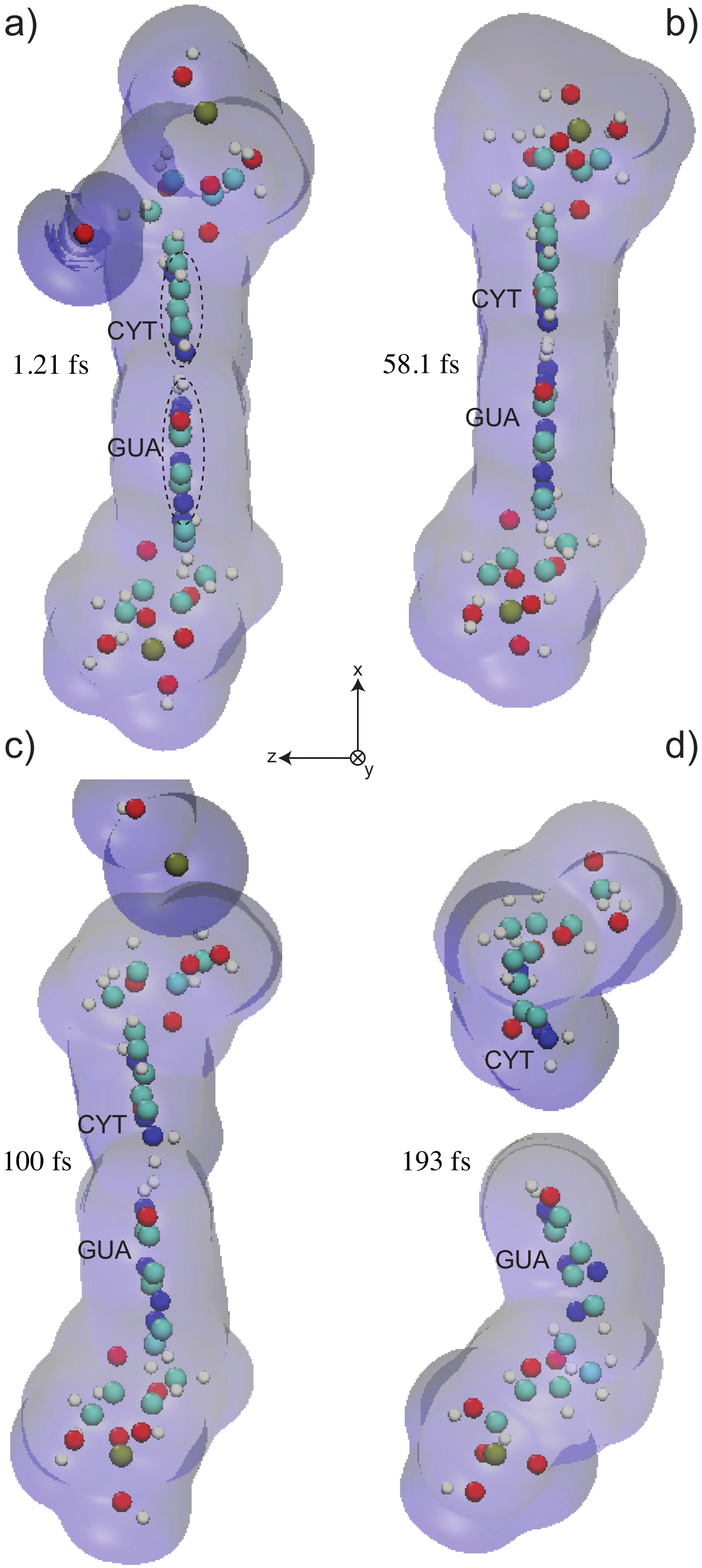}
\caption{System snapshots for some important stages during the proton-bound DNA  upper collision:  a) the proton just passing through the bp, b) the proton leaves the bp taking a fraction of the system charge (electron capture process), c) beginning of the dissociation process, and d) the bp dissociates with into five fragments. Animation with the whole process can be found in supplementary information.}
\label{diss-bound-P}
\end{figure}

Finally, Fig. \ref{diss-bound-P} shows snapshots of the proton DNA collision when the projectile impinges the phosphorus atom located on the cytosine side. This is a head-on collision where the proton transfers 61.8 eV to the bp. Unlike the two previous configurations, this is violent impact against the phosphorus atom so that 30.74 eV are immediately transferred to this ion, almost 50 \% of the total energy transfer. This is an energy transfer high enough to even break the P-O and P=O bonds, which requires about 3.67 eV and 6.03 eV for breaking, respectively. At $\sim$2 fs and $\sim$10 fs, O1P and O2P atoms are ejected from the impacted phosphate group. At $\sim$100 fs, even the phosphorus atom is emitted, together with another oxygen atom and a proton. According to Fig. \ref{BLc}, hydrogen bonds are dissociated, despite that the impact is relatively far from this region. Only the O5' atom remains linked to the sugar in this sugar-phosphate group. In addition, the sugar is dissociated from the cytosine base, yet the process is slower than hydrogen-bonds dissociation. Again, the sugar-guanine bond survive to the proton impact. C5'-C4' and C5-O5' in both sugars oscillates but do not break so the deoxyribose integrity seems to be preserved.

\section{Discussion}
According to this TD-DFT picture, DNA is initially ionized by the proton, both by electron capture and direct ionization. Simultaneously,  the proton perturbs the whole system (see supplementary information with animations) and this perturbation together with the system ionization lead to  the target molecule dissociation. The linking of the DNA bp to their neighbor tends to keep the integrity of the DNA backbone when the proton impacts on the bp center so only bases are damaged. Base damage was present in all the situation studied in this work since the hydrogen bonds that keep bases together have a low binding energy. 

The energetic analysis of these results show that 17.9 eV transferred during a central impact would only produce base damage while keeping the DNA backbone intact. Current Monte Carlo-based approaches to study the early DNA damage induced by ionizing radiation suppose that an energy transfer to the phosphate-sugar group above a given threshold is enough to induce a single strand break. This threshold is commonly set to about 10 eV. Those approaches  also consider that only the DNA component directly impacted by the projectile is affected. Those components are commonly divided into the sugar-phosphate groups and the nitrogeneous bases. This work shows that an impact on the phosphate group with an energy transfer of about 60 eV would be enough to break the DNA back bone on the impacted side and to damage the bases as well while keeping intact the opposite sugar-phosphate group. Furthermore,  it is observed, as expected, that the proton-DNA collision is a many-body problem during which all atoms feel the proton impact and receive a fraction of the transferred energy. Thus, there is no passive DNA base pair component during the collision, unlike it is supposed on the vast majority of biophysical models based on Monte Carlo simulations mentioned just above.  A complete TD-DFT study of this problem could provide enough information in order to change the actual paradigm of such biophysical models. That is,  DNA damage probabilities could be determined as a function of the projectile impact parameter and several sites can be damage in a single impact, as the backbone and bases.

This study also provides insights on the time evolution of the DNA base pair after the proton impact. Dissociation times are consistent with those reported in the literature for large molecules. The linking of the base pair to its neighbors tends to delay the dissociation process, which is about 100 fs according to our results.

\section{Conclusions}
TD-DFT can be a useful tool to study early DNA damage induced by the impact of heavy charged particle. However, due to the enormous computing resources and time demanded even for a single DNA base pair, a systematic study of this process is difficult to accomplish. A complete study should include  the combination of many impact parameters, energies, and incidence directions of the projectile. Yet, some interesting features can be obtained even with a limited number of calculations. For instance, energy transfers required for DNA damage can be inferred. In addition, some light can be shed on the effect of base-pair linking to its neigbors and the importance of the impact site on the way DNA is damaged by the ionizing particle. A relatively high energy transferred during the impact of a proton on the phosphate-sugar group can induce a single strand break together with bases damage.  It seems that every backbone break is accompanied by base damage. This is an important point since current Monte Carlo-based approaches in computational radiobiology suppose that base damage is independent of backbone break. A central impact with energy transfer of less than 20 eV would not be enough to produce backbone breaks. Current biophysical models used to study the early DNA damage induced by charged particles can be improved with studies like this one.

\section*{Acknowledgements}

M.B. thanks the {\it Conselho Nacional para o Desenvolvimento Cient\'ifico e Tecnol\'ogico (CNPq)}, Brazil, for financing his research activities through the  project 306775/2015-8. A.R. and U.D.G. acknowledge financial support from the European Research Council(ERC-2015-AdG-694097), Grupos Consolidados (IT578-13), H2020-NMP-2014 project MOSTOPHOS (GA no. 646259), European Union\textquotesingle s H2020  programme under GA no.676580 (NOMAD) and COST Action MP1306 (EUSpec). The images of this work were made with VMD  software support. VMD is developed with NIH support by the Theoretical and Computational Biophysics group at the Beckman Institute, University of Illinois at Urbana-Champaign.

%\bibliographystyle{achemso}
%\bibliography{/Applications/DocumentosTex/tesisDr-1}
\providecommand*\mcitethebibliography{\thebibliography}
\csname @ifundefined\endcsname{endmcitethebibliography}
  {\let\endmcitethebibliography\endthebibliography}{}

%\begin{figure}[htbp]
%\begin{center}
%\includegraphics[width=\textwidth]{TOC-TJPC.pdf}
%\label{default}
%\end{center}
%\end{figure}

\end{document}